\documentclass[10pt,conference,onecolumn]{IEEEtran} 
\IEEEoverridecommandlockouts
\usepackage{cite}
\usepackage{amsmath,amssymb,amsfonts}
\usepackage{algorithmic}
\usepackage{graphicx}
\usepackage{textcomp}
\usepackage{xcolor}
\usepackage{siunitx}
\usepackage{booktabs}
\usepackage{float}
\usepackage{dblfloatfix}

\usepackage{epsfig}
\usepackage{caption}
\usepackage{algorithm}
\usepackage{algorithmic}
\usepackage[utf8]{inputenc}

\usepackage{booktabs,caption}
\usepackage[flushleft]{threeparttable}
\usepackage{multirow}
\usepackage{float}
\def\BibTeX{{\rm B\kern-.05em{\sc i\kern-.025em b}\kern-.08em
    T\kern-.1667em\lower.7ex\hbox{E}\kern-.125emX}}
    
\usepackage{tikz}
\usepackage{textcomp}
\usepackage{hyperref}
\usepackage{lipsum}

\newcommand\copyrighttext{%
  \footnotesize \textcopyright 2019 IEEE. This article has been accepted for publication in IEEE TVLSI, which may differ from the finally published version. Personal use of this material is permitted.
  Permission from IEEE must be obtained for all other uses, in any current or future
  media, including reprinting/republishing this material for advertising or promotional
  purposes, creating new collective works, for resale or redistribution to servers or
  lists, or reuse of any copyrighted component of this work in other works.
  DOI: 10.1109/TVLSI.2019.2926324}
\newcommand\copyrightnotice{%
\begin{tikzpicture}[remember picture,overlay]
\node[anchor=south,yshift=10pt] at (current page.south) {\fbox{\parbox{\dimexpr\textwidth-\fboxsep-\fboxrule\relax}{\copyrighttext}}};
\end{tikzpicture}%
}

\begin{document}

\title{Practical Approaches Towards Deep-Learning Based Cross-Device Power Side Channel Attack}

\author{Anupam Golder, Debayan Das, Josef Danial, Santosh Ghosh, Shreyas Sen, and Arijit Raychowdhury
        
\thanks{This work was supported in
part by the National Science Foundation (NSF) under Grant CNS 17-19235. A. Golder is also supported by the National Science Foundation (NSF) under Grant  CNS 16-24810 - Center for Advanced Electronics through Machine Learning (CAEML) and its industry members.}        
\thanks{A. Golder and A. Raychowdhury are with the School of Electrical and Computer Engineering, Georgia Institute of Technology, Atlanta, GA, 30332 USA (e-mail: anupamgolder@gatech.edu; arijit.raychowdhury@ece.gatech.edu).}

\thanks{D. Das, J. Danial and S. Sen are with the School of Electrical and Computer Engineering, Purdue University, West Lafayette, IN, 47907 USA   (e-mail: \{das60; jdanial; shreyas\}@purdue.edu).}
\thanks{S. Ghosh is with Intel Labs, Intel Corporation, Hilsboro, OR, USA.}
}

\maketitle
\copyrightnotice
\begin{abstract}
Power side-channel analysis (SCA) has been of immense interest to most embedded designers to evaluate the physical security of the system. This work presents profiling-based cross-device power SCA attacks using deep learning techniques on 8-bit AVR microcontroller devices running AES-128. Firstly, we show the practical issues that arise in these profiling-based cross-device attacks due to significant device-to-device variations. \textit{Secondly}, we show that utilizing Principal Component Analysis (PCA) based pre-processing and  multi-device training, a Multi-Layer Perceptron (MLP) based 256-class classifier can achieve an average accuracy of $99.43\%$ in recovering the first key byte from all the 30 devices in our data set, even in the presence of significant inter-device variations. Results show that the designed MLP with PCA-based pre-processing outperforms a Convolutional Neural Network (CNN) with 4-device training by $\sim20\%$ in terms of the average test accuracy of cross-device attack for the aligned traces captured using the ChipWhisperer hardware. \textit{Finally}, to extend the practicality of these cross-device attacks, another pre-processing step, namely, Dynamic Time Warping (DTW) has been utilized to remove any misalignment among the traces, before performing PCA. DTW along with PCA followed by the 256-class MLP classifier provides $\geq$10.97\% higher accuracy than the CNN based approach for cross-device attack even in the presence of up to 50 time-sample misalignments between the traces.

\end{abstract}


\section{Introduction}
\subsection{Motivation}

In today's world, to establish secure communication between two parties, use of cryptographic algorithms is commonplace. Although these mathematically-secure crypto algorithms cannot be broken by means of brute-force attack, there have been numerous accounts of breaking the secret key by utilizing side-channel information in the form of power consumption \cite{kocher1999differential}, electromagnetic radiation \cite{agrawal2002side,quisquater2001electromagnetic,gandolfi2001electromagnetic}, and optical  \cite{kuhn2002optical,loughry2002information}, or acoustic \cite{asonov2004keyboard} vibrations captured from hardware implementations of the algorithms.

In this article, we focus on power side-channel analysis (SCA) based attacks, and specifically on profiling-based attacks \cite{chari2002template}. In traditional non-profiled attacks, such as differential and correlation power analysis (DPA\cite{kocher1999differential}/CPA \cite{brier2004correlation}) attacks, the attacker gathers power traces from a target device and uses statistical techniques, such as the difference of mean traces or Pearson Correlation Coefficient to break the secret key. On the other hand, profiling-based attacks \cite{chari2002template},\cite{cagli2017convolutional},\cite{lerman2018template} assume the worst-case scenario from the perspective of the target crypto engine, where the adversary is assumed to possess an identical device to profile the leakage patterns for all possible combinations for a keybyte (profiling or training phase), and use this prior knowledge to identify the secret key of the victim's crypto engine (online or test phase). Such an attack has been demonstrated on a commercially available contactless smart card in \cite{oswald2011breaking}.

\begin{figure}[!t]
  \centering
  \includegraphics[width=0.48\textwidth]{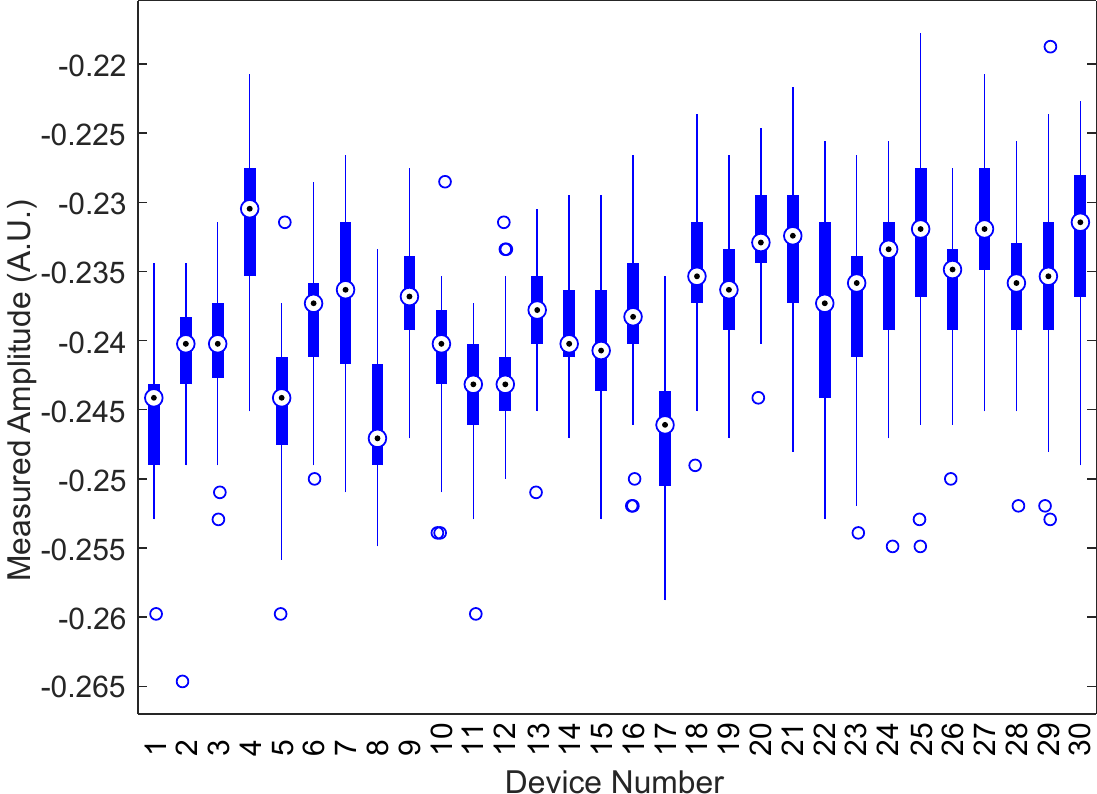}
  \caption{Box-and-whisker plots showing a distribution of 180 observations of sample \#96 of the power traces obtained from 30 identical microcontroller devices in two separate batches (CW308T-XMega, Batch1: Date Code: 1848, Lot Code: 0412 - Devices (1-20), Batch2: Date Code: 1830, Lot Code: 0350 - Devices (21-30)), running an AES encryption operation for a fixed plaintext and a fixed subkey $0x00$ for first keybyte. Amplitudes are reported in Arbitrary Unit (A.U.), as the ChipWhisperer capture platform does not report any unit for the measured power. The shift in medians and interquartile ranges can be attributed to manufacturing and packaging variations of different devices and circuit boards. Outliers, which fall outside the interquartile range for a specific distribution, are plotted in blue circles.}
  \label{boxplot_devices}
\end{figure}

Traditionally, profiling attacks are performed by generating templates (hence called $template \ attack$ \cite{chari2002template},\cite{choudary2018efficient},\cite{montminy2013improving},\cite{hanley2014empirical}) for different keys by utilizing a multi-variate Gaussian distribution approximation of the pre-identified points-of-interests (POIs). Recently, the hardware security research community has focused its attention to the machine learning (ML) based profiled attacks using Support Vector Machine (SVM) \cite{bartkewitz2012efficient},\cite{lerman2014power}, Random Forest (RF) \cite{lerman2014power} as well as deep learning based attacks\cite{cagli2017convolutional},\cite{benadjilastudy}, ~\cite{maghrebi2016breaking}, ~\cite{martinasek2016profiling}). The advantage of deep learning based attacks is that not only do they perform as good as the template-based attacks, they do not require extensive statistical analysis to identify POIs. Moreover, as the number of dimensions increase, machine learning based attacks start to gain interest, because increase in number of non-informative points in POIs degrades performance of template attacks \cite{lerman2018template}. But these ML-based approaches mentioned above rely on the assumption that the leakage profile from the profiling and the target devices are similar. 

Recently, \cite{das2019dac} showed the first cross-device profiling-based attack using neural networks. However, the attack scenario was limited to a small sample of devices taken from the same batch. 

Hence, the test accuracy for cross-device attack obtained using a Multi-layer Perceptron (MLP) without any pre-processing was very optimistic. Also, in \cite{cryptoeprint:2019:054}, authors have evaluated a Secure RSA implementation using Deep Learning where the training, validation, and testing data were collected from three different smart cards only. 

In a practical attack scenario, the target device may come from a different batch other than the one on which the ML-based classifier has been trained. Hence, the inter-device variations may be significantly higher, making cross-device attacks more difficult. This paper thoroughly analyzes the performance of the MLP classifier by obtaining 30 devices from two different batches (Figure \ref{boxplot_devices}) and shows that using the MLP even when trained with 4 devices, the accuracy across the test devices may be as low as $\sim8\%$ for the devices from a different batch or from the same batch with high inter-device variation (refer to Fig. 5(d)). Although this work is an extension of \cite{das2019dac}, it presents a significant advancement in generalizing Cross-device attack by utilizing PCA-based pre-processing to project trace samples to their principal sub-space \cite{archambeau2006template}, along with the multi-device training of the MLP classifier, which together provide a drastic improvement in minimum value of test accuracy by boosting it up to $\sim90\%$ from $\sim8\%$. In addition, to extend and improve the practicality of this attack, we utilize Dynamic Time Warping (DTW), preceding the PCA stage, to align the traces in case of desynchronization. Hence, DTW and PCA followed by the MLP classifier outperforms the CNN-based classifier by $>10\%$ in the test accuracy, even in presence of up to 50 time sample misalignments among the traces. This extended paper includes the following key improvements over \cite{das2019dac}: (a) A more detailed analysis on the performance of Deep Learning techniques in Cross-Device attack scenario using traces collected from 30 devices, (b) A plausible explanation of improvement in test accuracy using multi-device training (Section III.C), (c) Use of pre-processing to improve test accuracy instead of using multiple traces as presented in \cite{das2019dac} (Sections IV and V), and (d) A comparative analysis with CNN (Sections III.E and V) 

Figure \ref{boxplot_devices} shows a box-and-whisker plot of distributions of the measured amplitude of power consumption at a specific point in time from 30 different but identical microcontroller based AES-128 crypto engines. Sample point \#96 has been chosen using a feature selection method typically used for template attacks (Section III), namely, Difference of Means (DOM). Note how the medians and interquartile ranges vary from one device to another, and some outliers can be observed even for a particular device. It is clear from this figure that we need to validate our assumptions regarding identical leakage patterns from identical but different devices.

\subsection{Contribution}
Specific contributions of this paper are:
\begin{itemize}
\item This work analyzes the practical feasibility of using a 256-class classifier using MLP and Convolutional Neural Network (CNN) without any pre-processing to implement a cross-device attack on an 8-bit microcontroller based SCA platform named ChipWhisperer \cite{o2014chipwhisperer} (Figure \ref{setup}), and demonstrates how multi-device training improves the test accuracy of cross-device attack (Section III) with a plausible explanation.

\item Using Principal Component Analysis (PCA) \cite{jolliffe2011principal} based projection of raw traces to their principal subspaces along with multi-device training, we show that the accuracy of cross-device attack for all the test sets improve significantly (minimum test accuracy increases by $>10\times$), with $\geq$ 89.21\% accuracy across all the 30 devices (Section IV).

\item Finally, to enhance the practicality of the proposed attack, we consider the misalignment between traces due to untimely triggering of the capture device (usually an oscilloscope). To resolve this issue, we demonstrate that Dynamic Time Warping (DTW) \cite{muller2007dynamic} based pre-processing preceding PCA is a feasible solution, achieving $\geq$ 10.97\% higher accuracy compared to a CNN-based approach (Section V) even in the presence of up to 50 time sample misalignments between traces for a cross-device attack scenario.

\end{itemize}

\subsection{Paper Organization}
In this article, we use the following conventions: uppercase boldface italics for matrices, lowercase boldface italics for vectors, uppercase and lowercase italics for scalars, and uppercase italics within curly braces for sets.

The remainder of the paper is organized as follows. Section 2 presents the related works in the field of profiling-based attacks and summarizes the existing works on machine-learning based side-channel attacks. In Section 3, two Deep Neural Networks, MLP and CNN architectures have been presented for a single-trace attack. Also, their limitations in cross-device attack scenario are presented. Section 4 proposes a Principal Component Analysis based pre-processing step to project raw traces to their principal subspace, and thus effectively increase the cross-device attack accuracy. Section 5 investigates a more practical scenario considering misalignment between traces which can occur during trace capture, and we utilize a DTW based pre-processing to re-align traces, to allow subsequent PCA and MLP-classifier to work properly, and compare the performance of such an approach with that of a CNN. In Section 6, we present a relative timing performance comparison among different Deep Learning techniques, and between Deep Learning technique and CPA. Finally, in Section 7, we summarize the findings and conclude the paper.

\section{Background and Literature Review}

In the past couple of years, several ML techniques have been investigated, including but not limited to support vector machines (SVM) \cite{lerman2014power,bartkewitz2012efficient}, random forests (RF) \cite{lerman2014power}. More recently, the signal processing community as well as the hardware security researchers have started exploring the field of Deep Neural Networks (DNNs) \cite{cagli2017convolutional,gilmore2015neural}.
\subsection{Related Works}
Deep Learning based profiling attacks have been successful even in the presence of masking \cite{gilmore2015neural}, and jitter/misalignment \cite{cagli2017convolutional} based countermeasures. But other than \cite{das2019dac} and ~\cite{cryptoeprint:2019:054}, none of the articles proposing ML-based attacks investigated a practical scenario where the attack is actually performed on a device other than the one used in training phase.

\begin{table*}[!b]
\centering
\caption{Literature Review for Profiled-Attack Scenario}
\begin{threeparttable}

\begin{tabular}{|c|c|c|} 
\hline
\textbf{Profiled-Attack Scenario}    & \textbf{Method}             & \textbf{Corresponding Articles}  \\ 
\hline
\multirow{6}{*}{Same-device Attack}  & Template Attack & ~\cite{chari2002template}, ~\cite{rechberger2004practical}, ~\cite{oswald2007template}                                 \\ 
\cline{2-3}
                                     & Support Vector Machine      &   ~\cite{bartkewitz2012efficient}, ~\cite{lerman2014power}, ~\cite{lerman2015machine}, ~\cite{heuser2012intelligent}                             \\ 
\cline{2-3}
                                     & Random Forest               &  ~\cite{lerman2014power}                                \\ 
\cline{2-3}
                                     & Self-Organizing Map         &  ~\cite{lerman2014power}                                \\ 
\cline{2-3}
                                     & Time Series Approach        &  ~\cite{lerman2013time}                                \\ 
\cline{2-3}
                                     & Neural Networks             & ~\cite{cagli2017convolutional}, ~\cite{gilmore2015neural}, ~\cite{maghrebi2016breaking},~\cite{benadjilastudy}, ~\cite{martinasek2016profiling},~\cite{martinasek2013optimization}                                 \\ 
\hline
\multirow{2}{*}{Cross-device Attack} & Template Attack &  ~\cite{montminy2013improving},~\cite{choudary2018efficient},~\cite{renauld2011formal},~\cite{hanley2014empirical}                                \\ 
\cline{2-3}
                                     & Neural Networks             & ~\cite{das2019dac},~\cite{cryptoeprint:2019:054}, This Work                                  \\
\hline
\end{tabular}
\centering 

\end{threeparttable}
\end{table*}

Table 1 summarizes the related works. As can be seen, most of the template attacks were evaluated on the same device, whereas only in a few cases, (\cite{montminy2013improving},~\cite{choudary2018efficient},~\cite{renauld2011formal},~\cite{hanley2014empirical}) attacks were performed on a different device. This article significantly improves on \cite{das2019dac} to present a generalized Deep Learning based cross-device SCA attack on 30 different devices, utilizing DTW and PCA along with the multi-device training.

There are two types of classification strategies for ML-based classifiers, one is based on the Hamming-Weight (HW) model (9-class classification) and the other is the identity (ID) model (256-class classification). When the ID model is used, the attack typically requires a single trace from the target device. Thus, these attacks are powerful in the event the adversary has limited time and opportunity to gather such traces due to re-keying after each session. This article utilizes the ID model to leverage the advantages of such an attack model.

\begin{figure}
  \centering
  \includegraphics[width=0.45\textwidth]{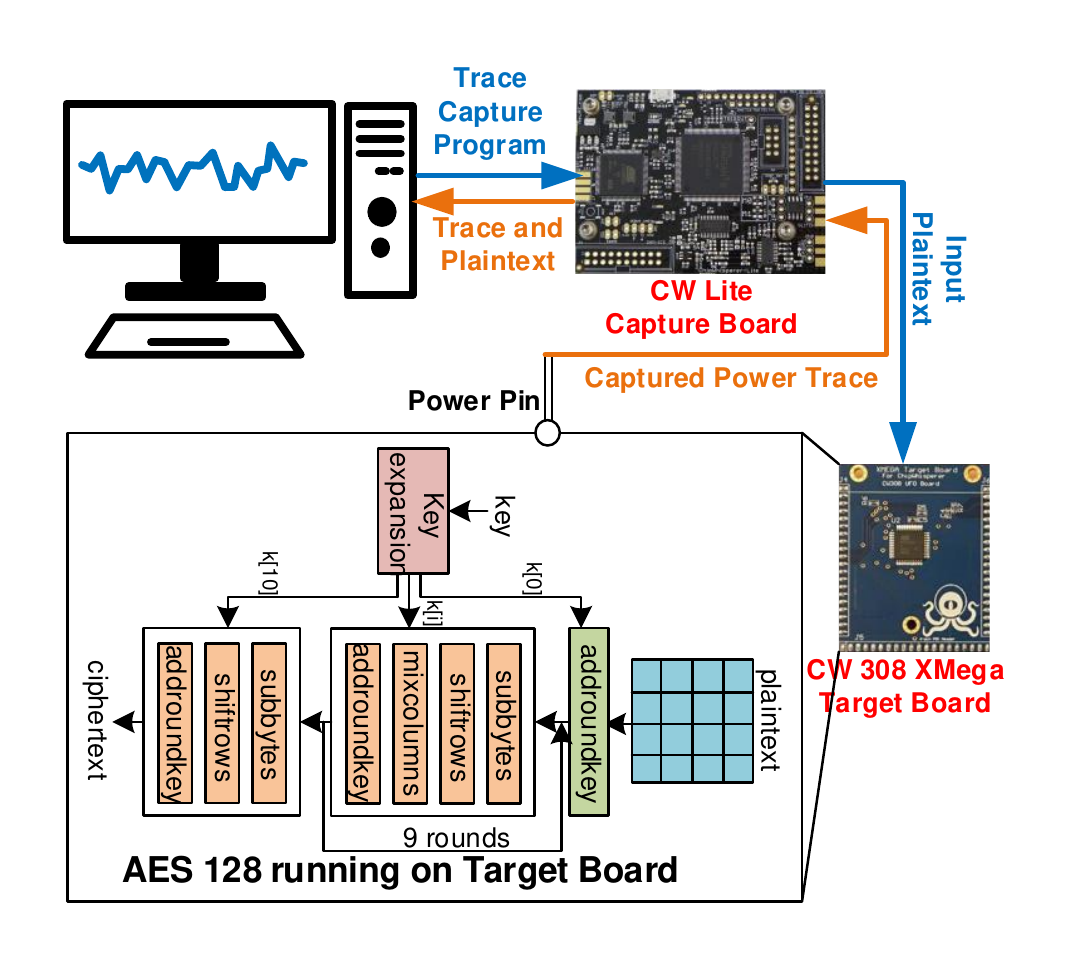}
  \caption{ChipWhisperer platform for capturing and recording power traces from an AVR XMega microcontroller running AES-128 encryption algorithm. This platform supports programming the microcontroller to execute AES-128, modification of keys and plaintexts, and capturing of power traces with a perfectly triggered circuit to ensure alignment between traces.}
  \label{setup}
\end{figure}

Most of the previous ML-based attacks were evaluated using the DPAv2 \cite{telecomdpa} and DPAv4 \cite{telecom2014dpa} contest data sets, or recently published ASCAD \cite{benadjilastudy} database. To the best of our knowledge, both data sets consist of traces captured from one single device, which are not suitable for current work. Hence, we collected new traces from 30 different 8-bit AVR microcontrollers running the AES-128 algorithm using the ChipWhisperer platform \cite{o2014chipwhisperer} (Figure \ref{setup}).  Although 8-bit microcontrollers are becoming less preferred for encryption engines nowadays, recent body of work (\cite{choudary2018efficient}, \cite{benadjilastudy}, \cite{cryptoeprint:2018:1023}, \cite{yang2018convolutional}, \cite{picek2018improving}) investigated performance of Profiled SCA attack using datasets gathered from 8-bit microcontrollers.
\subsection{Background of Neural Networks}
One of the most widely used Neural Networks is Multi-Layer Perceptron which has an input layer, one or more hidden layers, and an output layer. Each layer has trainable weights and biases, and produces an output based on its inputs coming from the preceding layer, weights connected to those inputs, and a non-linear activation function, such as, Rectified Linear Units (ReLU), sigmoid, or $tanh$. Choice of the number of layers, number of neurons in each layer, and the activation function/s together create the network architecture and define the functions that can be approximated by this model. In a classification problem, the last layer is a classification layer followed by a softmax layer. More complex architectures are possible such as CNN, which uses so-called convolutional layers that act as filters and slide over the preceding layer by 1 or more units (called stride), subsampling layers such as maxpooling or average-pooling to reduce number of dimensions, and the fully connected layers. The gathered data is usually divided into three distinct sets: training set, as the name implies, to train the network, validation set to validate the performance of trained network on previously unseen data by usually keeping a part of training set separate from those actually used in training, and a test set, which is used to finally test the performance, i.e., prediction or classification accuracy. Typically, during the training phase, weights and biases of neurons are tuned iteratively over several epochs to minimize a categorical cross-entropy loss \cite{lecun2005loss} function using a form of Stochastic Gradient Descent (SGD) optimizer \cite{kingma2014adam}. The training phase can be conducted in mini-batches where the inputs for the network are continuously drawn from the complete set of training samples. On top of these, use of batch normalization, dropout layer and $L_2$ regularization are typical ways to address the problem of overfitting (when the model works well for the training set, but performs poorly on the unseen test set) to provide a better generalization on the test data. All the parameters that define the architecture of neural networks, and dictate the training phase are called hyper-parameters.

Note that, in contrast to template based attack, where a template preparation method can be exported across platforms, the trained neural network architecture may be vastly different depending on the implementations. In other words, one particular neural network architecture for a specific attack scenario and a specific hardware implementation may not be the best suited one for a different platform.

\subsection{Overview of Principal Component Analysis}
Principal Component Analysis (PCA) \cite{jolliffe2011principal} is a well-known dimensionality reduction technique, and has been proven to be successful for time-series data. Say, we have a $M\times N$ matrix of traces, where each trace forms a row. In the following matrix representations, $t_{ij}$ denotes sample $j$ of trace $i$, and each column vector, $\boldsymbol{t_{j}}$ (each having a dimension of $M\times 1$) represents data for $j$-th dimension.

\[
\boldsymbol{Traces} = \begin{bmatrix} 
    \boldsymbol{trace_{1}} \\
    \boldsymbol{trace_{2}}\\
    \vdots \\
    \boldsymbol{trace_{M}}
    \end{bmatrix}
    = \begin{bmatrix} 
    t_{11} & t_{12} & \dots & t_{1N}\\
    t_{21} & t_{22} & \dots & t_{2N}\\
    \vdots & \vdots & \ddots & \vdots\\
    t_{M1} & t_{M2} & \dots & t_{MN}
    \end{bmatrix}
\]
\[
= \begin{bmatrix} 
    \boldsymbol{t_{1}} & \boldsymbol{t_{2}} & \dots & \boldsymbol{t_{N}}
    \end{bmatrix}
\]
Then, we subtract the mean from each of the dimensions, to obtain a new matrix, $\boldsymbol{Traces_{adjust}}$.
\[
\boldsymbol{Traces_{adjust}}= 
\]
\[
\begin{bmatrix} 
    \boldsymbol{t_{1}}- mean(\boldsymbol{t_{1}}) & \boldsymbol{t_{2}}- mean(\boldsymbol{t_{2}}) & \dots & \boldsymbol{t_{N}} -mean(\boldsymbol{t_{N}})
    \end{bmatrix}
\]
Next, the covariance (a measure of variation of each of the dimensions from their individual means with respect to each other; helpful for very high dimensional data) matrix for the traces is computed as follows:
\[
Covariance \ matrix, \boldsymbol{C} = cov(\boldsymbol{Traces}) 
\]
\[
= cov(\begin{bmatrix} 
    \boldsymbol{t_{1}} & \boldsymbol{t_{2}} & \dots & \boldsymbol{t_{N}}
    \end{bmatrix})
\]
\[
    = \begin{bmatrix} 
    cov(\boldsymbol{t_{1},t_{1}}) & cov(\boldsymbol{t_{1},t_{2}}) & \dots & cov(\boldsymbol{t_{1},t_{N}})\\
    cov(\boldsymbol{t_{2},t_{1}}) & cov(\boldsymbol{t_{2},t_{2}}) & \dots & cov(\boldsymbol{t_{2},t_{N}})\\
    \vdots & \vdots & \ddots & \vdots\\
    cov(\boldsymbol{t_{N},t_{1}}) & cov(\boldsymbol{t_{N},t_{2}}) & \dots & cov(\boldsymbol{t_{N},t_{N}})\\
    \end{bmatrix}
\]

Then, we calculate the unit eigenvector matrix of the covariance matrix, $\boldsymbol{V}$, and the diagonal eigenvalue matrix, $\boldsymbol{D}$. These unit eigenvectors are orthogonal to each other.
\[
\boldsymbol{V} = \begin{bmatrix} 
    \boldsymbol{v_{1}} & \boldsymbol{v_{2}} & \dots & \boldsymbol{v_{N}}
    \end{bmatrix}
\]
Eigenvectors with the highest eigenvalues are the most significant principal components of the data set. Ordering eigenvalues from highest to lowest arranges the components according to the order of importance. If eigenvalues are very small, principal components corresponding to those of lesser importance can be discarded with negligible information loss. Here, the original data dimension was $N$, and we have correspondingly $N$ eigenvalues and $N$ eigenvectors. Choosing the first $p$ eigenvalues after arranging them in descending order of eigenvalues, reduces the dimension for the dataset to $p$. Then, we obtain a modified eigenvector matrix, $\boldsymbol{V_m}$ keeping the first $p$ eigenvectors in that matrix. Finally a new dataset, $\boldsymbol{Traces_m}$ is derived using:
$$\boldsymbol{Traces_m} = (\boldsymbol{V_m}'\times \boldsymbol{Traces_{adjust}}')' \ [eqn. 1]$$
where $\boldsymbol{'}$ denotes a matrix transpose operation.\\
PCA has been studied extensively (\cite{archambeau2006template},\cite{batina2012getting},\cite{choudary2014efficient},\cite{cagli2015enhancing}) in SCA context, but rarely so when it comes to Deep Learning techniques. In contrast to Template Attacks, where dimension reduction is necessary, Deep Learning can handle large dimensions, and the benefit of PCA comes from projection of trace samples to their principal subspace, and not particularly from the pruning (Section IV.A).

\subsection{Dynamic Time Warping (DTW)}
\begin{figure}[!t]
  \centering
  \includegraphics[width=0.4\textwidth]{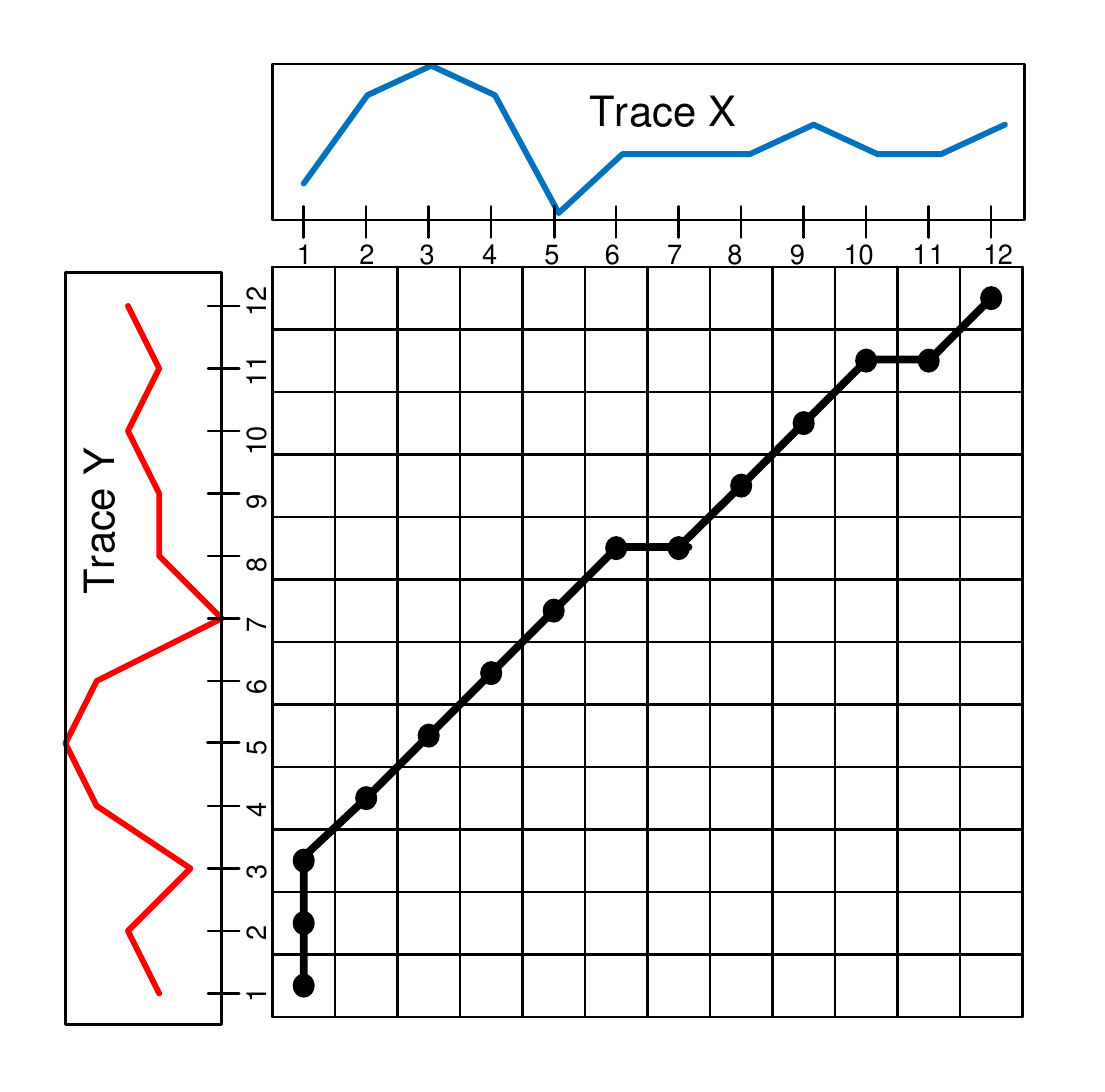}
  \caption{Overview of Dynamic Time Warping (DTW): Warp path for traces X and Y shows how both the traces are non-linearly resampled to ensure matching. For example, samples 1 and 2 in trace Y are absent in trace X. Hence, DTW samples X at the sample 1 multiple times. Then, samples 3-8 in the trace Y match with the samples 1-6 in trace X, and so on. In this way, two traces are realigned such that the absolute or Euclidean distance between them is minimized.}
  \label{warp_path}
\end{figure}

Misalignment between captured power traces can occur due to inaccurate triggering\cite{van2011improving}, or countermeasures adopted by device manufacturers such as frequency scaling\cite{baddam2007evaluation} or random insertion of dummy operations\cite{ambrose2007rijid}. As pointed out in \cite{van2011improving}, time series matching algorithms, such as Dynamic Time Warping (DTW), originally adopted for alignment in speech recognition systems\cite{sakoe1978dynamic}, can be beneficial to realign them in an attack scenario where only a limited number of traces from target device can be collected. But in contrast to \cite{van2011improving}, where DTW has been used as a pre-processing step before performing a non-profiled DPA attack or a profiled template attack\cite{chari2002template}, we propose to use it as a pre-processing step for a Neural Network Classifier.

DTW resamples two traces such that matching parts of them are located at same time index after the process, thus reducing distance (euclidean/absolute) between them to a minimum. The resampled indices in both traces form a $warp \ path$. Note that as DTW can align only one trace with respect to another, we need a $reference \  trace$ for realignment. Figure \ref{warp_path} illustrates a warp path $W$ = $\{z(k): 1\leq k \leq K\}$ which is set of $x(k)$ and $y(k)$ indices, $z(k)$ = $(x(k),y(k))$ of two traces $X$ and $Y$, under the following constraints: 
$$z(k+1) = (x(k+1),y(k+1))\\= \left\{
                \begin{array}{ll}
                  (x(k),y(k)+1), or\\
                  (x(k)+1,y(k)), or\\
                  (x(k)+1,y(k)+1))
                \end{array}
              \right.
$$
$\\and \ x(1)$ = $y(1)$ = $1$, $x(K)$= $y(K)$ = $T, T \leq K<2T\\$
$$where, T = number \ of \ samples \ in  \ X \ and \ Y\\$$ 

DTW algorithm tries to find a warp path $W$ that results in minimum cost $L$ given as:
$$L(X,Y) = \frac{1} {2T} \min_{W} \sum_{k=1}^{K} d(z(k))c(k)$$
$$where, d(z(k)) = |X(x(k))-Y(y(k))|$$
$$and \ c(k) = x(k) - x(k-1) + y(k) - y(k-1)$$

This process results in stretching both the traces (X and Y) by resampling them. The resulting traces become perfectly aligned.

\section{Single-Trace Cross-Device Power SCA using Neural Networks: Performance and Limitations}

In this section, we present the architectures of our designed Multi-Layer Perceptron and Convolutional Neural Network, along with the empirical choice of hyperparameters, and performance and limitations of these single-trace attacks without any pre-processing.

\subsection{Experimental Setup}
We performed our experiments on CW308T-XMega, a microcontroller based SCA platform from ChipWhisperer \cite{o2014chipwhisperer}. This microcontroller based target board houses an 8-bit Atmel AVR XMega128 microcontroller running software AES-128. To capture traces from the target device, ChipWhisperer provides a platform CW308T UFO board, and a capture setup CW Lite Capture using an on-board ADC. This setup allows to send program, plaintext and key to XMega Target board, and record captured traces directly from a personal computer (Figure \ref{setup}). The on-board XMega microcontroller operates at the frequency of 7.37 MHz, and CW Lite Capture hardware captures traces at 4 times of that frequency, at 29.48 MHz. Power consumption is measured by inserting a small resistor ($500 m\Omega$) in series with power supply and measuring voltage drop across it. As this measured voltage corresponds directly to instantaneous current drawn from the DC power supply, it can be treated as the power trace. We utilize a $chosen \ plaintext$ scenario for our attack, where we keep the plaintext fixed to a chosen value ($0x00...00$, i.e., all 16 bytes are 0s). We collect $10k$ traces ($3k$ time samples per trace) from each of the 30 devices by varying the first keybyte in all 256 possible combinations (from $0$ to $255$) maintaining a uniform distribution ($\frac{10k}{256} \approx 40$ traces for each possible keybyte value), and rest of the keybytes randomly. We attack only the first keybyte in all our results, but the attack can be carried out for all other keybytes in the same manner. Neural network models were implemented using MATLAB and Python, using the keras\cite{chollet2018keras} library with tensorflow\cite{abadi2016tensorflow} as backend.

\subsection{Architecture of Multi-layer Perceptron}

\begin{figure}
  \centering
  \includegraphics[width=0.45\textwidth]{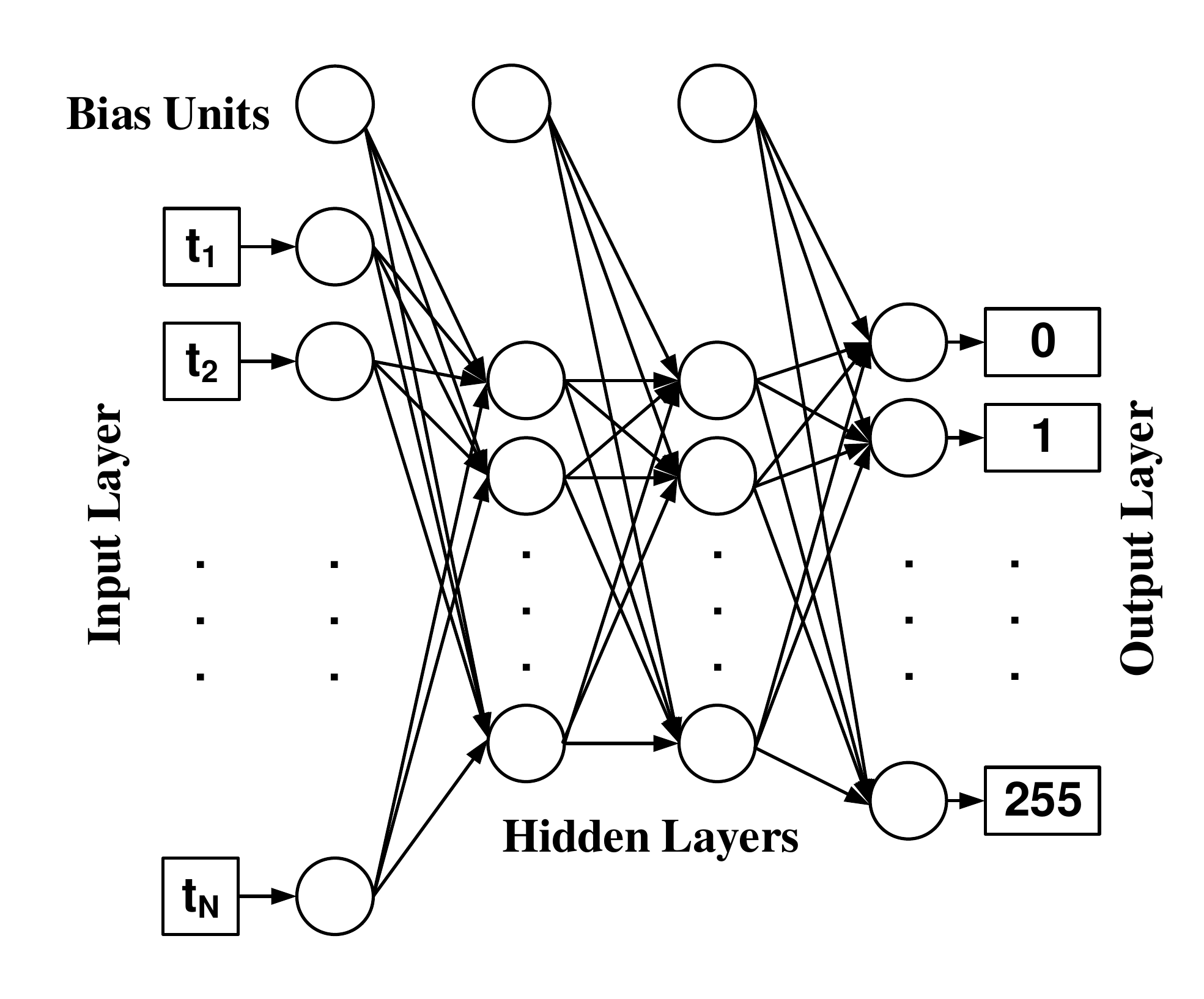}
  \caption{ Architecture of the proposed Multi-Layer Perceptron for Cross-device Side Channel Attack. The input layer consists of N = 3000 neurons. The $1^{st}$ fully-connected (FC) hidden layer consists of 100 hidden neurons, followed by Batch Normalization, Rectified Linear Unit (ReLU) activation, and a dropout layer. The $2^{nd}$ hidden layer is similar without the dropout layer. Finally, the output layer has 256 neurons for predicting the correct key byte utilizing the softmax function.}
  \label{MLP_architecture}
\end{figure}

Figure~\ref{MLP_architecture} shows the architecture of our designed Multi-Layer Perceptron (MLP) for the 256-class classification,which is similar to the one presented in \cite{das2019dac}. The size of the raw or processed traces determine the size of the input layer of the MLP. 2 fully-connected layers each consisting of 100 neurons form the hidden layers. Activation function for the layers is chosen as ReLU. A dropout layer after first hidden layer having a percentage dropout of 10\% has been chosen to aid generalization. $L_2$ regularization seemed to have little effect. We keep it fixed at $10^{-4}$ for all the iterations. Also, the mini batch size is kept at 256, the network has been evaluated after being trained for 100 epochs. It was observed that higher number of hidden neurons and layers led to overfitting to the training data, resulting in relatively poor performance in test set. Also, choice of the batch size is a critical issue. It was observed that small batch sizes led to high test accuracy. To optimize the network, we train the MLP with $8k$ training traces for single-device training, and optimize the performance by validating against a test set of the remaining $2k$ traces from the same device.

\begin{table*}
\caption{Cross-Device attack Performance of Deep Learning-based Methods for different training Scenarios*} 
\centering 
\begin{threeparttable}

\begin{tabular}{c c c c c c c c c c} 
\hline\hline 
 $\boldsymbol{Number \ of \  }$ & \multicolumn{9}{c}{$\boldsymbol{Test \ Accuracy \  (\%)}$}\\
$\boldsymbol{Training \  Devices}$ &\multicolumn{3}{c}{$\boldsymbol{MLP}$}&\multicolumn{3}{c}{$\boldsymbol{PCA-MLP}$}&\multicolumn{3}{c}{$\boldsymbol{CNN}$}\\
  & Average & Maximum & Minimum & Average & Maximum & Minimum & Average & Maximum & Minimum
\\ [0.5ex]
\hline 
\\
1 &61.98  &98.70  &2.95 &  90.09 &99.94 &53.18   &29.97 &44.86 &10.09 \\ 
2 &79.14  &99.92  &4.47 &  96.65 &\textbf{99.99} &71.28   & 47.75 & 74.42 & 21.27\\
3 &90.76  &99.93  &8.93 &  99.37 &\textbf{99.99} &\textbf{90.82}   &78.69 & 98.93 & 51.15\\
4 &91.72  &99.95  &8.02 &  \textbf{99.43} &\textbf{99.99} &89.21   &80.39 &94.63 &60.08 \\  [1ex]
\hline 
\end{tabular}

\centering 
\begin{tablenotes}
\item \textit{*Does not include Test Accuracy for Devices used in Training Set} 
\end{tablenotes}
\end{threeparttable}

\end{table*}

\subsection{Performance of MLP}
\begin{figure*}[!b]
  \centering
  \includegraphics[width=1\textwidth]{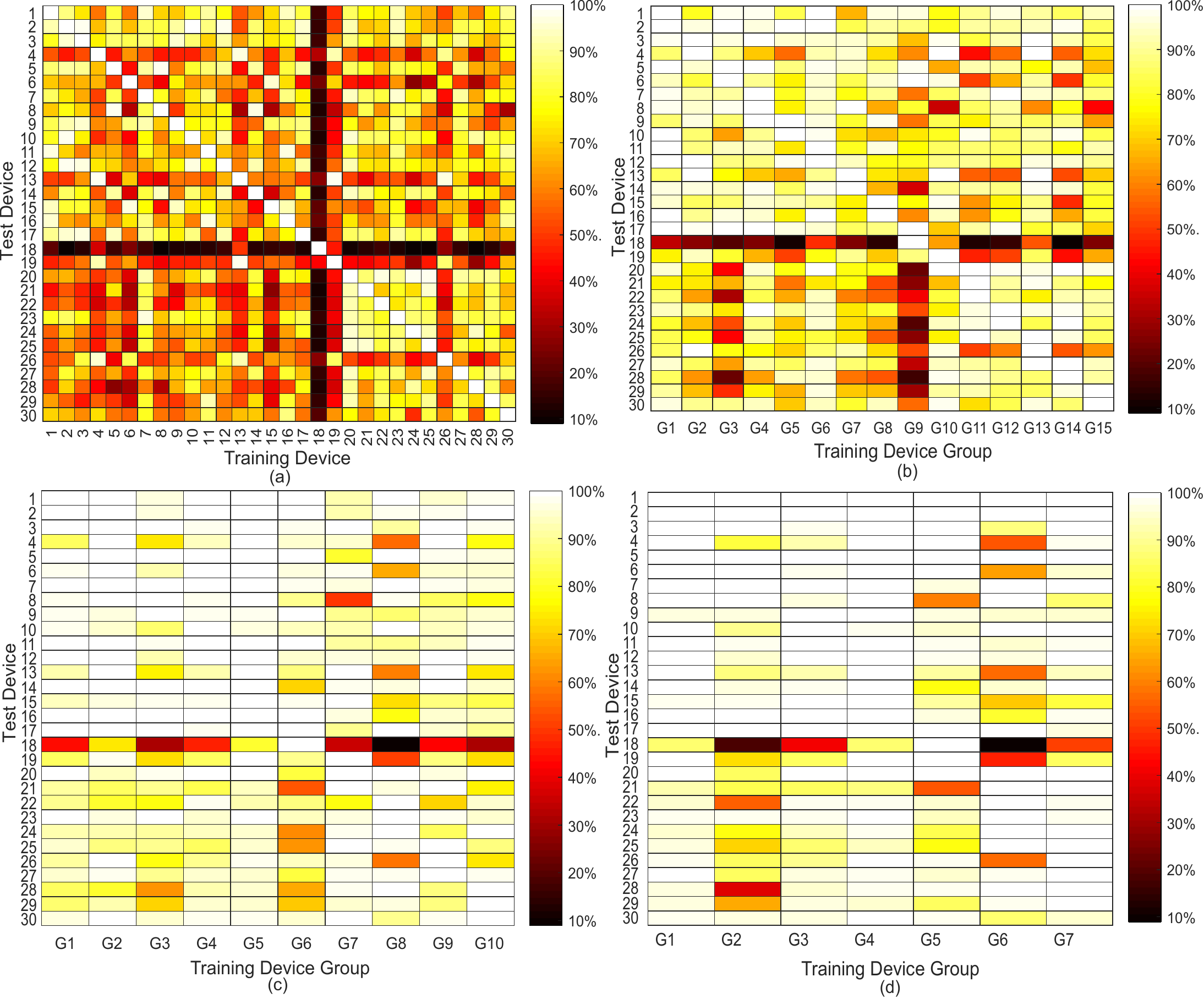}
  \caption{Test Accuracy of MLP classifier when number of devices in training set is (a) one (b) two (c) three (d) four. As can be seen from (a), although the test accuracy for the $same$ $device$ $attack$ is very high it varies widely in case of a $cross$ $device$ $attack$. Particularly, notice how Device \#18 has 100\% test accuracy when the classifier is trained with that device, but a very low accuracy when the neural network is trained with other devices. Figure (c-d) illustrates improvement in test accuracy with increase in number of training devices.}
  \label{heatmap_MLP}
\end{figure*}

\begin{figure}[!t]
  \centering
  \includegraphics[width=0.45\textwidth]{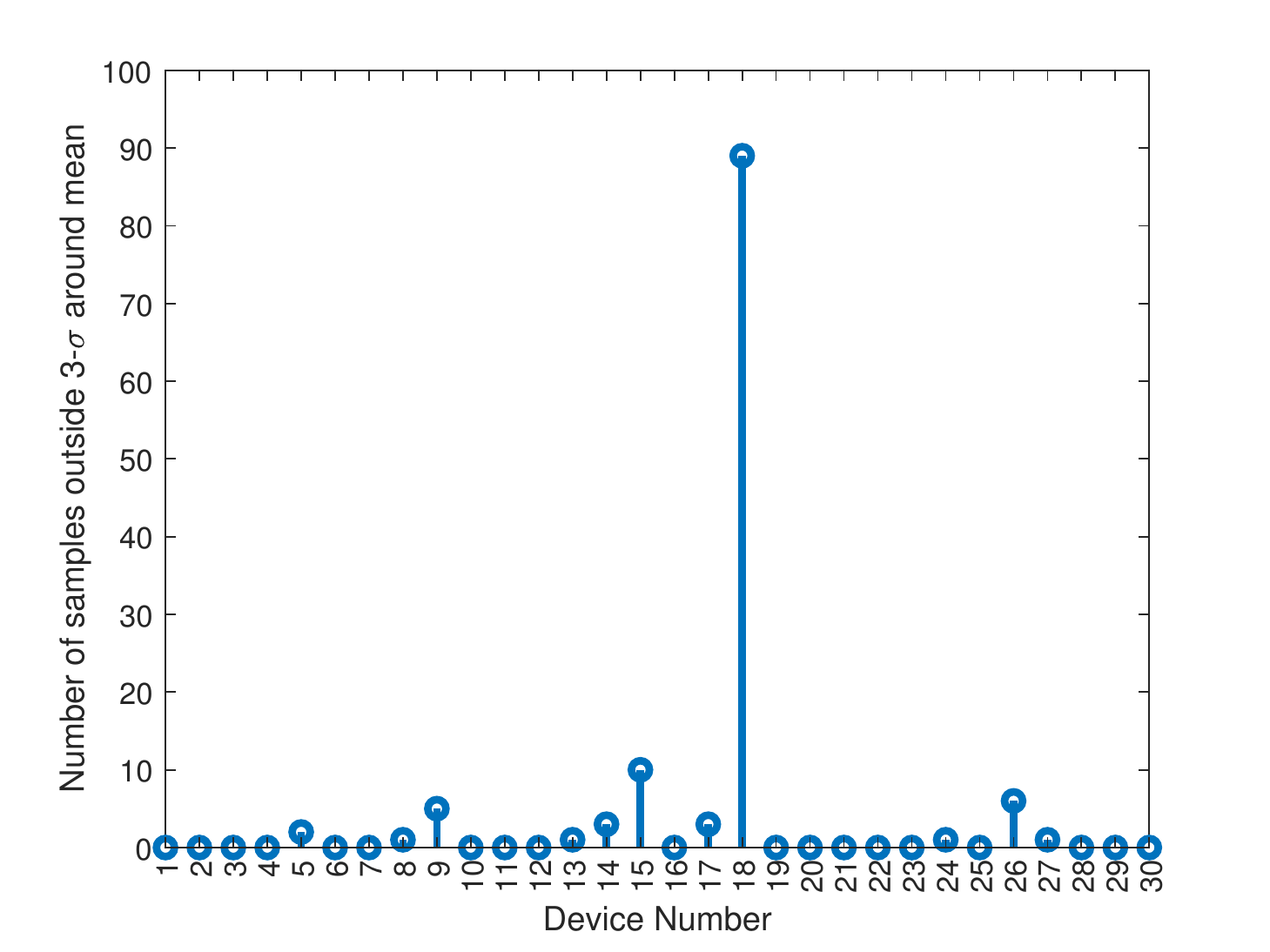}
  \caption{Rationale behind Device 18 being an outlier in Figure \ref{heatmap_MLP}(a): $\sim 90$  samples of averaged trace for Device 18 fall outside the $3\sigma$ range around the mean for averaged traces across all devices.}
  \label{dev18}
\end{figure}

In this section, we evaluate the performance of MLP. The problem with one device training is that the neural network overfits to that device-specific leakage and cannot generalize to new data set from a different device. As shown in Figure \ref{heatmap_MLP}(a), although same-device attack performance of MLP is very high ($\geq$ $99.99\%$), cross-device attack performance is relatively poor, averaging at $61.98\%$ across the 30 devices after training for 100 epochs (Table II). Also, we note from Figure \ref{heatmap_MLP}(a) that Device 18 is clearly an outlier, although we cannot see much deviation for Device 18 in Figure \ref{boxplot_devices}. In Figure \ref{boxplot_devices}, we only observed the distribution for one specific dimension of a 3000-dimensional trace. That is why, it does not give us the whole picture. To find out why Device 18 is an outlier in this experiment, the average for each time sample of all 8k traces (with 3000 samples each) has been calculated from the training set for all 30 devices in our dataset. Then the mean $\mu$, and standard deviation $\sigma$ have been calculated for each time sample of the averaged traces from all 30 devices. Assuming an approximately Gaussian distribution, 99.7\% of the time samples of the averaged traces for each device should fall within 3 standard deviation around the mean ($\mu+3\sigma$ and $\mu-3\sigma$). Then, the number of samples (out of 3000) of averaged traces for each device which fall outside this range has been counted. In Figure \ref{dev18}, the result has been illustrated, where Device 18 is certainly an outlier, which explains why we obtained poor test accuracy for Device 18 when the neural network was trained with traces from other devices, and vice versa.\\

\begin{figure}[!b]
  \centering
  \includegraphics[width=0.45\textwidth]{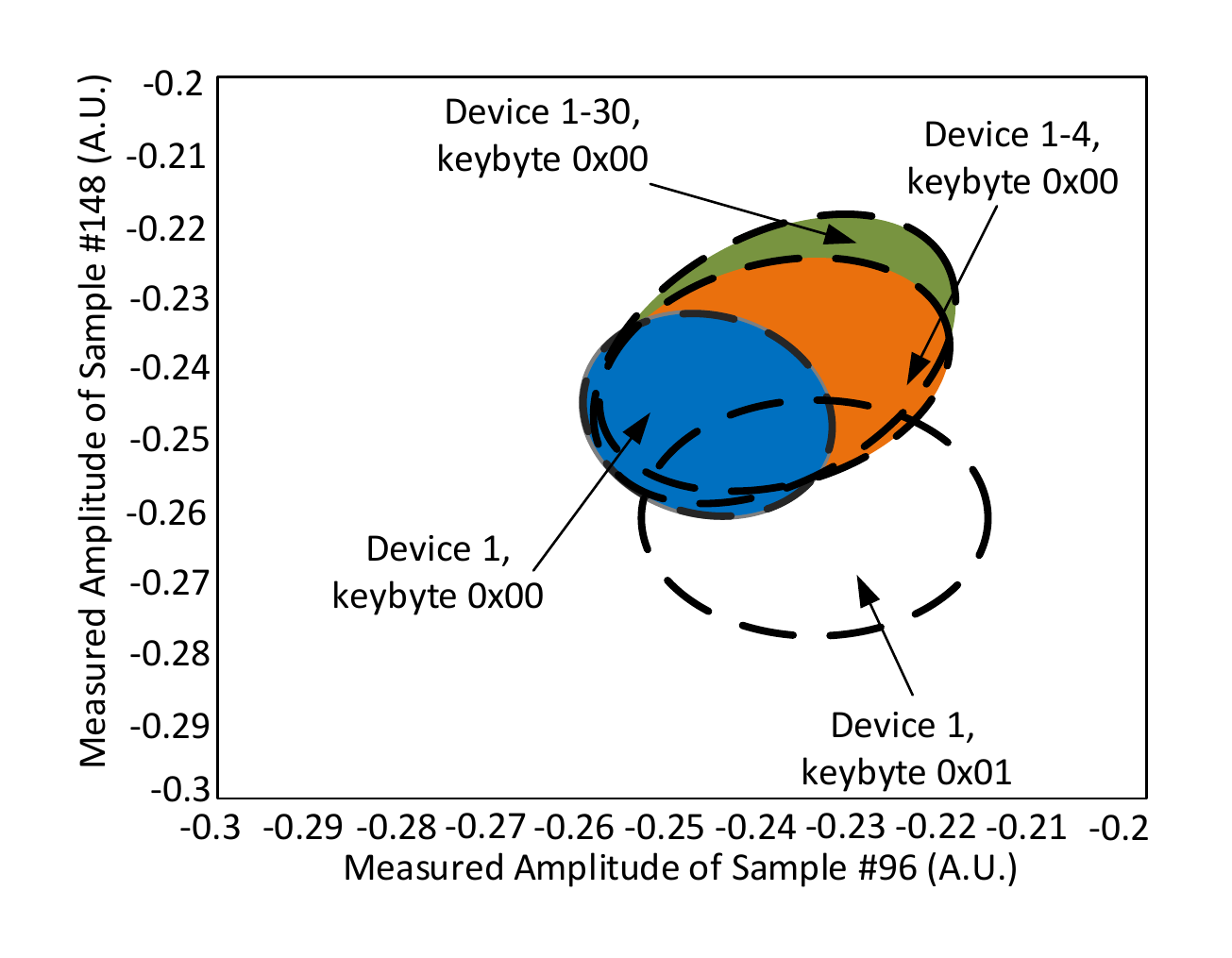}
  \caption{Boundary of Bivariate Probability Density Function (PDF) for different scenarios. Two of the most prominent leakage samples from the raw power traces are chosen as the variables for constructing the bivariate distribution. Note that, the PDFs are different for different keybyte values. Also, as the number of devices increase from 1 to 4, the sample PDF for a specific keybyte value (0x00) approximates the total PDF of the 30 devices more accurately.}
  \label{4dev_exp}
\end{figure}

To eliminate the problem of low test accuracy of MLP with single-device training, we perform an empirical evaluation of the multi-device training method \cite{das2019dac}, so that the neural network learns from the device manufacturing and packaging variations and generalizes to test traces from a new device. It should be noted that although this would strengthen the adversary, the attack becomes more difficult to implement, as it assumes more control on the part of the adversary. Nevertheless, such a multi-device training improves the average test accuracy to as high as $91.72\%$ as illustrated in Figure \ref{heatmap_MLP}(b-d) and Table II. Note that, we merge the data sets for different devices before multi-device training, and use $20k$, $30k$,and $40k$ traces in training and validation set for the 2-device, 3-device, and 4-device training respectively. To incorporate cross-validation, but at the same time bearing in mind the training time required for an exhaustive one, we construct several training device groups and use them for all subsequent analysis. Training groups for multi-device training are formed using the formulation: $ \{ G_j(i) \} =\{ D(k), D(k+1), ..., D(k+j-1)\mid  k = (i-1)j+1 \ and \  k+j-1 \leq 30 \}$, where $\{G_j(i)\}$ is the $i-th$ group used in $j-$ device training, and $D(k)$ is the $k-th$ device in our dataset. Also note that, further increasing the number of traces (or more devices) in training set did not improve test accuracy.

Although the improvements with multi-device training seem intuitive, in this article, we further investigate the factor that could have led to this. In this regard, we seek inspiration from a POI selection method, namely, Difference of Means \cite{chari2002template},\cite{choudary2018efficient}, typically used in template attacks. Using the sum of the absolute value of pairwise differences between mean of traces for different keys (keeping plaintext fixed), we observed that the samples \#96 and \#148 are two such POIs. To achieve a more accurate result, template attacks typically identify 10-20 such POIs, and calculate a probability density function (PDF) using a Multi-variate Gaussian Distribution approximation. Although such a high-dimensional template would give us more accurate results, we concentrated on a simple bivariate analysis (as it can be visualized in 2D) to observe if a pattern could be identified from the distribution. The probability density function for multivariate normal distribution of k-variables is given as:
$$f_{\boldsymbol{x}} = \frac {exp(-1/2({\boldsymbol{x}}-{\boldsymbol{\mu}})^T\ {\boldsymbol{\Sigma}}^{-1}({\boldsymbol{x}}-{\boldsymbol{\mu}}))}{\sqrt{(2\pi)^k |{\boldsymbol{\Sigma}}|}} $$
where, ${\boldsymbol{\Sigma}} = Covaraince \  Matrix $ and ${\boldsymbol{\mu}} = Mean \  Vector$

Covariance matrix and mean vector for the bivariate normal joint density function have been calculated from experimentally measured data. Figure \ref{4dev_exp} shows the boundary regions of PDFs. It can be seen that data from different keys lead to different distributions for the same device. Also, traces from a single device do not span the whole distribution of 30 devices, but with 4 devices, most of it can be spanned. This is plausibly why our trained MLP model achieved a high average test accuracy with 4-device training.

\subsection{Limitation of only MLP based attack}
Notice from Table II that the minimum test accuracy does not improve much with increasing number of training devices, but the average test accuracy improves significantly. This motivates us to investigate further into pre-processing techniques such as PCA to improve the minimum test accuracy (Section IV), and utilize the multi-device training to prevent overfitting and aid generalization.

\subsection{Architecture and Performance of Convolutional Neural Network (CNN)}
\begin{figure}[!b]
  \centering
  \includegraphics[width=0.4\textwidth]{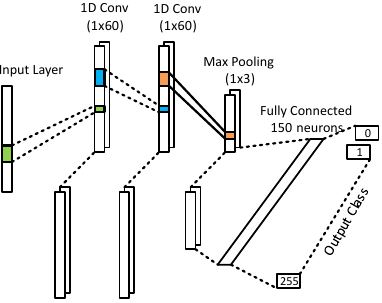}
  \caption{Architecture of 1-D CNN: Samples of raw traces are directly fed to the input layer. Two 1-D convolutional layers act as filters and extract high level features from raw traces. Next, a Max Pooling layer reduces the number of dimensions by subsampling. The outputs of Max Pooling layer are flattened and provided as inputs to a Fully Connected Layer which connects the to final classification Layer.}
  \label{CNN_architecture}
\end{figure}

We designed a one-dimensional (1-D) Convolutional Neural Network (CNN) architecture (Figure \ref{CNN_architecture}), as shown in the recent works \cite{cagli2017convolutional}, \cite{benadjilastudy} to have very high accuracy in the presence of countermeasures. CNN architecture presented here is different from those works, as it is designed for our new data set. We gathered inspiration from VGG-Net \cite{simonyan2014very}, but used a much shallower network. Such a choice again depends on the dataset. 

\begin{figure*}[!t]
  \centering
  \includegraphics[width=1\textwidth]{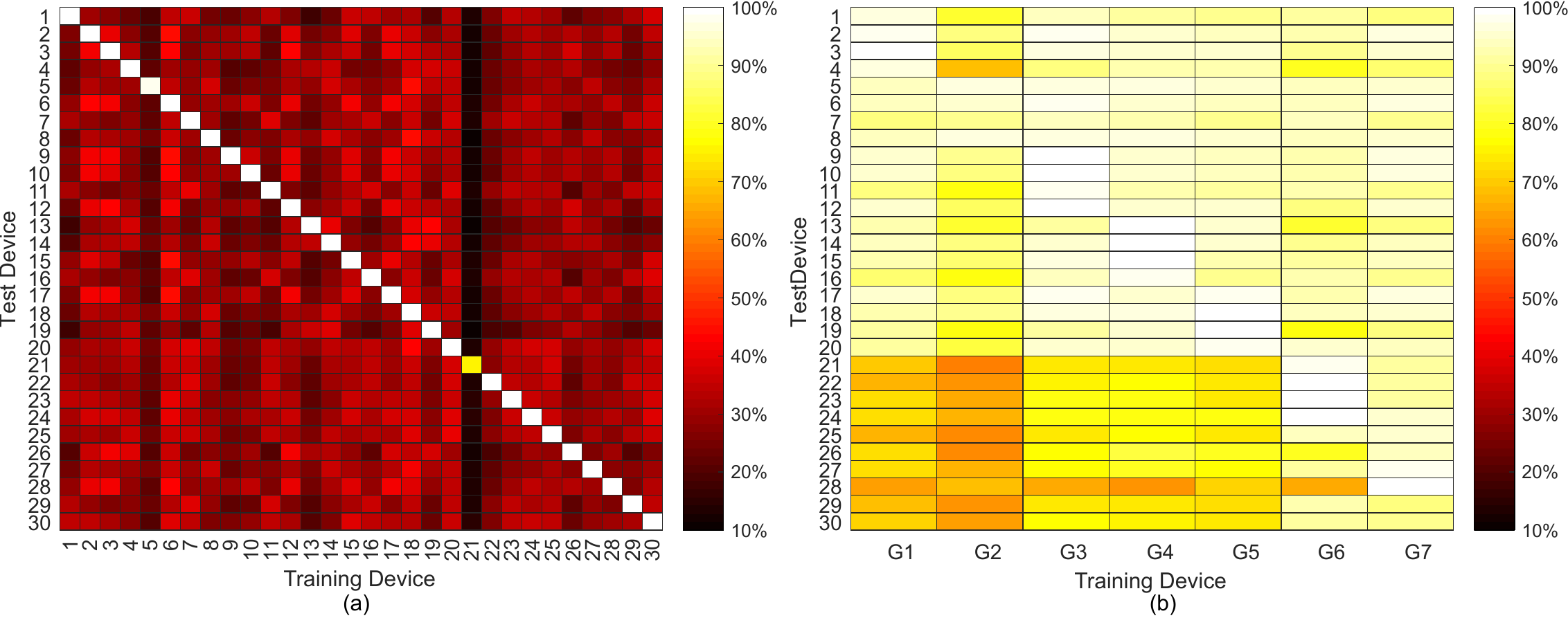}
  \caption{Test Accuracy of CNN classifier when number of devices in training set is (a) one (b) four. As can be seen from (a), although the test accuracy for the $same$ $device$ $attack$ is very high, it is extremely low in the case of a $cross$ $device$ $attack$. Figure (b) illustrates improvement in test accuracy with increase in number of training devices, but compared to MLP, the average accuracy still remains lower.}
  \label{heatmap_CNN}
\end{figure*}

We kept the same input layer as in the MLP, i.e., 3000 neurons. In this architecture, the first convolutional layer has 70 filters, each having a kernel size of 60, with a default stride of 1, and with ReLU activation function. The second convolutional layer is exactly the same. Then a maxpooling layer with a pool size of 3 was used. A subsequent layer flattens the output of maxpooling layer, followed by a fully connected layer with 150 neurons, a Batch Normalization Layer, and finally the classification layer. A dropout factor of 20\% after the flatten layer, and 10\% after the Batch Normalization layer reduced overfitting. The parameters have been optimized by validating on a test set from the same-device, similar to the approach in the case of the MLP architecture. Note that, CNNs require a larger data set to generalize well to new data, and we augmented the available data set by introducing normally distributed noise (with $10^{-10}$ standard deviation) to trace samples, and thus increasing number of training traces to $60k$ in all training scenarios. We evaluate the performance of CNN for single-device training and multi-device training after 20 epochs. Figure \ref{heatmap_CNN}(a) illustrates result of single-device training for CNN, and we can see that, this CNN network achieves very high test accuracy for traces from the same-device, but cross-device attack accuracy is low. Figure \ref{heatmap_CNN}(b) shows the improvement in cross-device attack accuracy with multi-device training. However, comparing figure \ref{heatmap_MLP} to figure \ref{heatmap_CNN}, we can observe that MLP outperforms the CNN in the cross-device attack performance (although comparable for the same device attack) for both single-device and 4-device training. 

Table II summarizes the comparison of cross-device performance between the MLP and the CNN models with multi-device training. Compared to the MLP, CNN provides better minimum accuracy, but lower average accuracy across the 30 devices taken from 2 different batches. Hence, we choose MLP as our desired classifier for the next sections, and develop strategies to achieve high accuracies for all the 30 devices.

\section{Performance of MLP with PCA-based Pre-processing (PCA-MLP)}
\begin{figure*}[!t]
  \centering
  \includegraphics[width=1\textwidth]{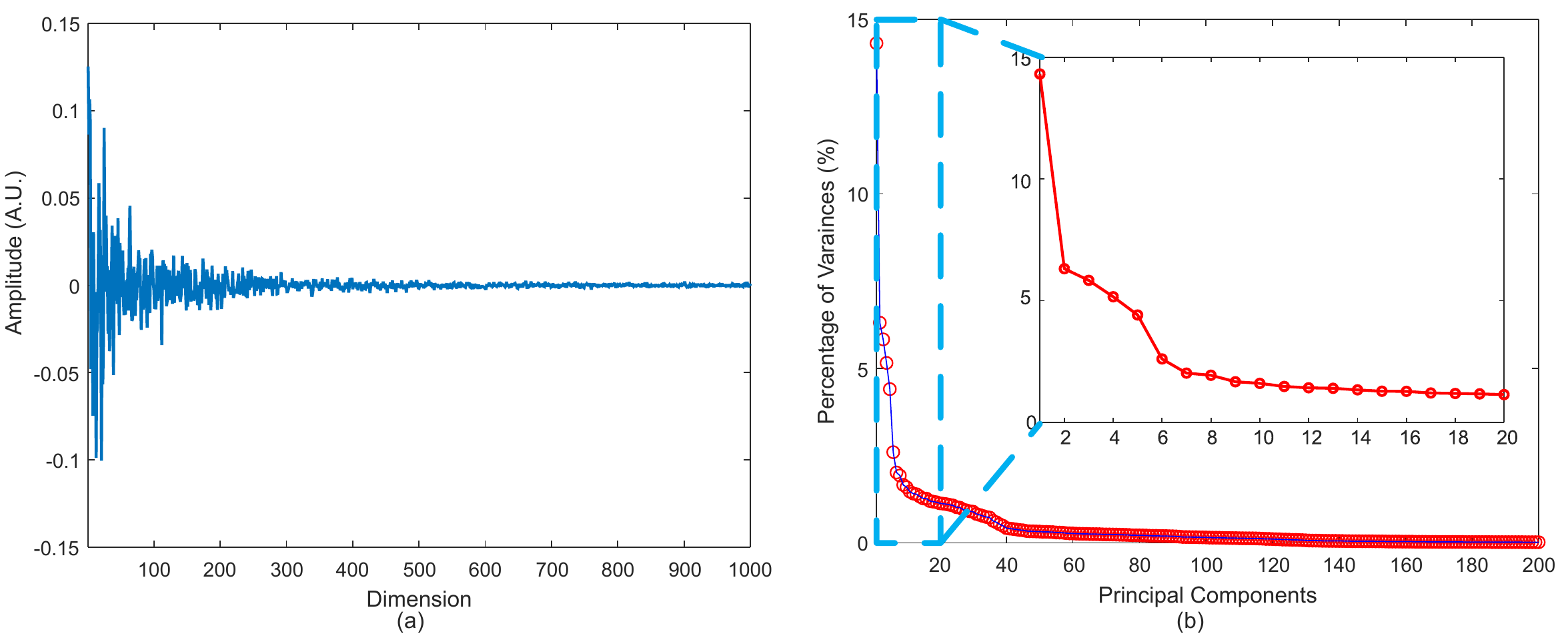}
  \caption{Principal Component Analysis of Raw Traces: (a) Transformed Trace after PCA (first 1000 Time Samples). Note that, amplitudes are lower with higher dimensions (b) Contribution of each principal component as percentage of total variances (first 200 principal components). The zoomed-in region corresponds to first 20 samples. Note that, first 60 samples contribute the most to the total variance. }
  \label{PCA_components}
\end{figure*}

In this section, we evaluate the effect of Principal Component Analysis (PCA) \cite{jolliffe2011principal} as a pre-processing step to enhance the performance of cross-device attacks. Figure~\ref{PCA_components}(a) shows the amplitude of the features extracted from raw traces using $eqn. 1$. 
Notice that, the eigenvectors with higher eigenvalues point to the direction of higher variance in data. As a result, in the transformed trace (Figure \ref{PCA_components}(a)), the samples on the left have higher amplitudes, and as we proceed to the right, the amplitudes decrease. From Figure~\ref{PCA_components}(b), we can see that, the first 60 time samples contribute the most to the total variance. Also, it has been observed that, for this data set, $99\%$ of total variance is contributed by the first 370 Principal Components.

\begin{figure*}
  \centering
  \includegraphics[width=0.97\textwidth]{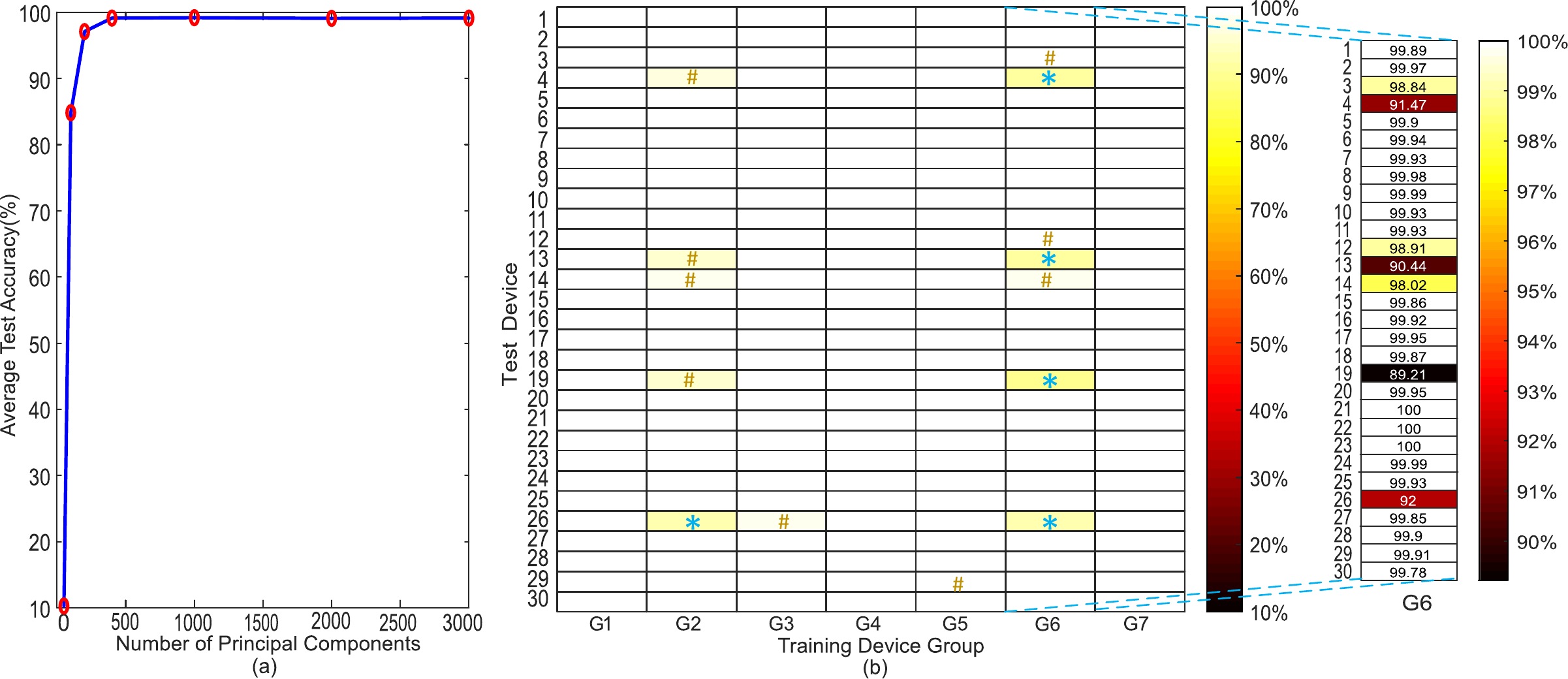}
  \caption{Performance of PCA-MLP: (a) Average Test Accuracy(\%) vs. number of principal components used (for Training Device Group G1). Test Accuracy does not improve on removing the principal components with lower contribution in the percentage variance. Hence, we chose to include all the 3000 principal components in the PCA-MLP model. (b) Test Accuracy after training PCA-MLP with 4 devices shows drastic improvement in the minimum cross-device accuracy. Note that, '*' symbol has been used to highlight the cases with 89-94\% test accuracy, and '\#' symbol for 95-98\% test accuracy.}
  \label{PCA_MLP_four}
\end{figure*}

\begin{figure*}[!t]
  \centering
  \includegraphics[width=0.95\textwidth]{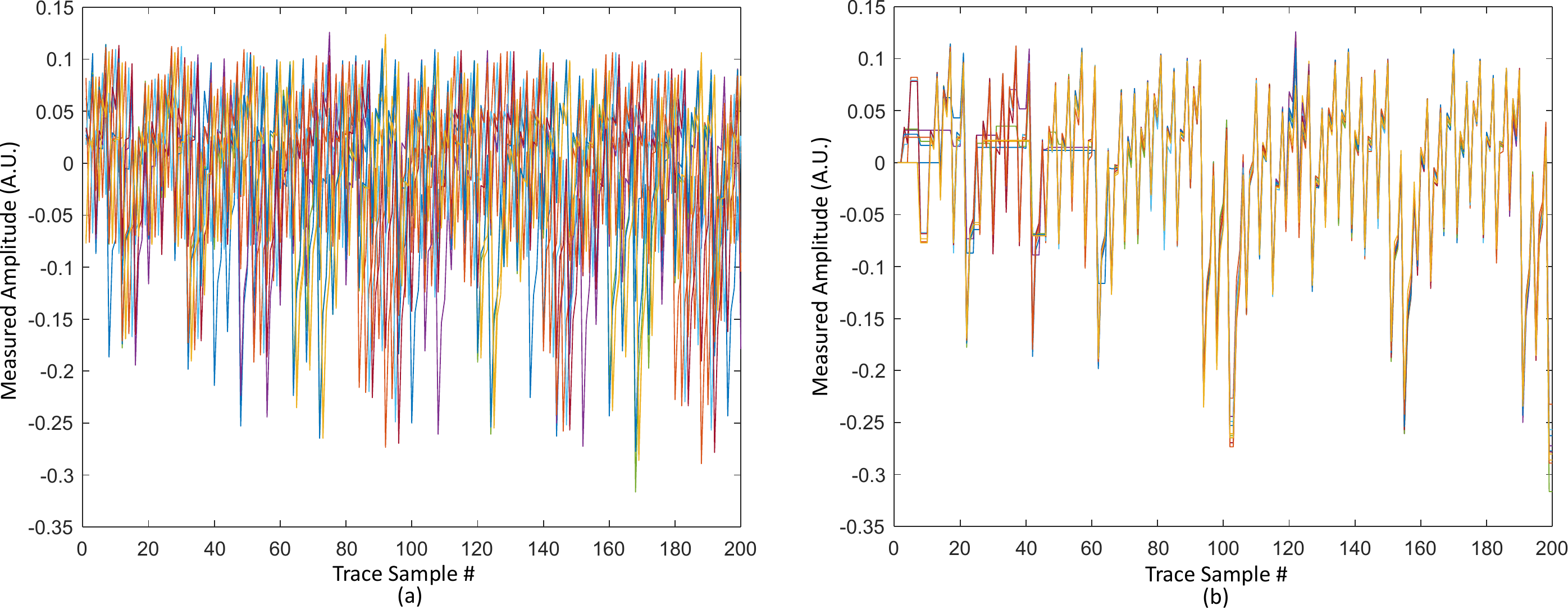}
  \caption{Ten Traces superimposed on each other (first 200 Samples shown) when they are (a) misaligned randomly up to 50 time samples (b) realigned using DTW}
  \label{trace_DTW}
\end{figure*}

\subsection{Performance of PCA-MLP}
As seen earlier, MLP-based classifier without pre-processing achieves good average accuracy ($>$ $90\%$) with multi-device training, but the minimum accuracy is as low as $\sim8\%$ (Figure~\ref{heatmap_MLP}, and Table II). In this section, we show that with PCA as a pre-processing step, the performance of the neural network classifier is substantially improved due to projection of raw trace samples to their principal subspace. For a principal subspace, coordinate axes are ordered in a way that they point to the direction of maximal variance in data \cite{archambeau2006template}.

Figure~\ref{PCA_MLP_four}(b) shows that both the minimum and the average test accuracy of PCA-MLP improve significantly with 4-device training compared to only MLP (Figure \ref{heatmap_MLP}(d)). Also, as summarized in Table II, it can be seen that the average as well as the minimum test accuracy improve going from single-device to multi-device training. Since with 4 devices, the PCA-MLP model achieves the best average accuracy, we chose 4 as the number of devices for multi-device training. It should be noted that the same number of training traces were used to train this model as reported in Section III. Figure \ref{PCA_MLP_four}(a) illustrates that reducing number of dimensions did not improve average test accuracy in our case. Hence, we used all the 3000 principal components in the analysis presented in this section, although higher dimensions have less informative features.

\subsection{Limitation of PCA-MLP}
One inherent assumption in the previous sections was that the traces were all perfectly aligned (as they were collected using the ChipWhisperer capture setup), which may not always be the case in a practical scenario due to faulty triggering. Figure~\ref{trace_DTW}(a) shows the traces when they are misaligned. The limitation of PCA and MLP is that the traces need to be perfectly aligned. This motivates us to investigate ways to re-align traces so that the benefits of the high classification accuracy for PCA-MLP can be utilized. \cite{cagli2017convolutional} proposed using CNN in case of misaligned traces, but as PCA-MLP showed better cross-device attack performance, we chose to use PCA-MLP, and to account for the misalignment, we adopted Dynamic Time Warping (DTW) as a pre-processing step (Section V).

\section{Dynamic Time Warping as a pre-processing for misaligned traces}
 
In this section, we show how Dynamic Time Warping (DTW) can be used to realign misaligned traces due to a fault in triggering. As ChipWhisperer platform perfectly synchronizes each capture event, the traces obtained from CW308T-XMega target board are perfectly aligned with each other. To simulate the event of trace misalignment, we artificially create up to 50 time-sample misalignments to evaluate the performance of proposed DTW-PCA-MLP architecture. Misalignment has been created by shifting traces by a random number of samples, in the same manner presented in \cite{benadjilastudy}. Figure~\ref{trace_DTW}(a) shows such a collection of 10 traces superimposed on each other. Notice that, the traces are randomly misaligned up to 50 samples. We chose 5 devices from our set of 30 devices, and created misaligned dataset $M1-M5$. Then, we evaluated performance of our proposed approach on a $cross-device \ attack$ to show its effectiveness by training on 4 devices from the misaligned data set and testing on the other one, and performing cross-validation for all the possible combinations.

\begin{algorithm}[!t]
\caption{Algorithm for DTW based Trace realignment} 
{\textbf{Input:} Misaligned Trace Matrix ($\boldsymbol{Tr}_m^{M \times N}$), Reference Trace ($\boldsymbol{tr}_{ref}^{1 \times N}$) \\
{\textbf{Output:} Aligned Trace Matrix ($\boldsymbol{Tr}_a^{M \times W_M}$), Modified Reference trace ($\boldsymbol{tr}_{refm}^{1 \times W_M}$) } 

\begin{algorithmic}[h!]  

\STATE $ W_0 \gets N$ 
\FOR{$i \gets 1$ to $M$}
	
    	\STATE $[\boldsymbol{x}_i^{1 \times W_i}, \boldsymbol{y}_i^{1 \times W_i}]$ $\gets$ $DTW(\boldsymbol{{Tr}}_{m}^{M \times N}(i,:),\boldsymbol{tr}_{ref}^{1 \times W_{i-1}})$
    	\IF{$i > 1$}
    	     \STATE $\boldsymbol{{Tr}}_{a}^{M \times W_i} \gets \boldsymbol{{Tr}}_{a}^{M \times W_{i-1}}(1:i-1,\boldsymbol{y}_i^{1 \times W_i})$
        \ENDIF
        
        \STATE $\boldsymbol{Tr}_{a}^{M \times W_i}(i,:) \gets \boldsymbol{{Tr}}_{m}^{M \times N}(i,\boldsymbol{x}_i^{1 \times W_i})$
        \STATE $\boldsymbol{tr}_{refm}^{1 \times W_i} \gets \boldsymbol{tr}_{refm}^{1 \times W_{i-1}}(1,\boldsymbol{y}_i^{1 \times W_i})$
   
\ENDFOR 
\STATE return $\boldsymbol{Tr}_a^{M \times W_M}$, $\boldsymbol{tr}_{refm}^{1 \times W_M}$
\end{algorithmic}
}
\end{algorithm}

\textcolor{red}{\begin{table*}[!h]
\caption{Performance Comparison of DTW-PCA-MLP with CNN for misaligned traces} 
\centering
\begin{tabular}{c c c c c c} 
\hline\hline
$Training \ Set$ & $Test \ Set$ & \multicolumn{4}{c}{$Test Accuracy(\%)$}\\& & DTW-PCA-MLP & CNN & DTW-CNN & DTW-PCA-CNN\\
\hline 
\\
M1-M4       &M5 &99.80 &87.05 &88.91 &89.63\\ 
M1-M3,M5    &M4 &99.71 &88.37 &95.53 &93.22\\
M1-M3,M4-M5 &M3 &99.69 &88.72 &92.64&90.16\\
M1,M3-M5    &M2 &99.94 &78.98 &92.41 &95.66\\
M2-M5       &M1 &98.86 &80.61 &92.44 &95.40\\  [1ex]
\hline 
\end{tabular}
\label{tab:PPer}
\end{table*}}

\begin{table*}[b!]
\caption{Qualitative Comparison between Different Deep Learning based Attack Methods} 
\centering
\begin{threeparttable}

\begin{tabular}{|c|c|c|c|c|c|} 
\cline{3-6}
\multicolumn{1}{l}{}                                                                                       &                   & \multicolumn{4}{c|}{\textbf{Average Test Accuracy of Different Methods*}}  \\ 
\cline{3-6}
\multicolumn{1}{l}{}                                                                                       &                   & \textbf{MLP}      & \textbf{PCA-MLP}   & \textbf{DTW-PCA-MLP} & \textbf{CNN}                         \\ 
\hline
\multirow{2}{*}{\textbf{Same-Device Attack}}                                                                        & \textbf{Aligned Traces}    & Very High     & Very High      & Very High        & Very High                        \\ 
\cline{2-6}
                                                                                                           & \textbf{Misaligned Traces} & Very Low & Very Low  & Very High        & Very High                        \\ 
\hline
\multirow{2}{*}{\begin{tabular}[c]{@{}c@{}}\textbf{Cross-Device Attack}\\\textbf{with multi-device training~~}\end{tabular}} & \textbf{Aligned Traces}    & High     & Very High & Very High   & High                        \\ 
\cline{2-6}
                                                                                                           & \textbf{Misaligned Traces} & Very Low & Very Low  & Very High   & High                        \\
\hline
\end{tabular}
\centering 
\begin{tablenotes}
\item \textit{*Evaluated on the same data set comprising of traces from 30 identical microcontroller devices} 
\end{tablenotes}
\end{threeparttable}

\end{table*}

\subsection{Implementation of DTW-PCA-MLP}
As mentioned in Section II, Dynamic Time Warping (DTW) method requires a reference trace to realign another trace. Obtaining such a reference trace is possible for an adversary from the device/s he has in his possession. In our experiments, we use a reference trace that has a 3000 time-sample window, containing all relevant samples. The network has been trained on 4 devices and tested on one, as shown in Table III. We keep the Neural Network architecture for MLP the same as in previous sections, apart from a change in the size of input layer. With experiments, it has been observed that for realigned data set, dimensionality reduction after PCA improved test accuracy. This can be partly due to the fact that DTW resamples the traces to find the best match, and in doing that, some samples are copied multiple times. We empirically found that 600 features led to the best test accuracy. So, we modified the number of input neurons to 600, but the rest of the architecture remains the same. Algorithm for DTW based trace realignment is shown in Algorithm 1. Using this algorithm, traces have been realigned (Figure \ref{trace_DTW}(b)).

\subsection{Performance Comparison among different methods for misaligned traces}
After trace re-alignment using DTW, subsequent PCA based pre-processing and MLP based classifier resulted in $\geq 98.86\%$ test accuracy, compared to the CNN (which has been re-optimized to deal with the misalignment) with $\geq 78.98\%$ test accuracy. The minimum difference between test accuracy of DTW-PCA-MLP and CNN is $10.97\%$ for the test set M3. We expect this trend to continue if extended to all 30 devices of our data set, based on results presented in Table III. Consequently, based on the results, we can clearly see that DTW-PCA-MLP is a convenient method for cross-device profiled-attack for misaligned traces, not only because of high average test accuracy, but also due to its simpler architecture and less effort on the choice of hyperparameters and lower training time compared to CNN. For the sake of completeness, we also evaluate the performance of CNN with only DTW and both DTW and PCA based pre-processing. CNN is expected to benefit from realignment through DTW,  although CNN is believed to be able to handle misalignments intrinsically. To see if CNN benefits from DTW (and PCA) when traces are misaligned, we conducted the experiment and Table III summarizes the results. From Table III, we see that inclusion of DTW certainly helps improve test accuracy in all cases for CNN, but inclusion of PCA after DTW may improve test accuracy in some cases compared to DTW-CNN, but that does not generalize well across all sets.

\section{Relative Timing Performance Comparison between Different SCA approaches}
 In comparison to Profiled Attacks, Non-Profiled attacks, such as, CPA/DPA would require at least tens to thousands of traces to correctly identify the key. As shown in \cite{das2019dac}, even in low Signal-to-Noise (SNR) scenarios, the benefit of using Deep Learning techniques persists, as CPA requires $\sim10\times$ more traces for same level of accuracy.\\
Training time depends on network complexity, chosen hyper-parameters, size of the dataset, software platform (e.g. tensorflow, pytorch etc.), and last but not the least, the hardware used to train. We have conducted the training and testing phase for MLP and CNN on the same hardware platform, and we observed that for the same training set (4 training devices used), batch size (256), and number of epochs (100), CNN requires ~$6\times$ more time to achieve the same level of accuracy ($\geq99\%$) compared to MLP (453.4 seconds compared to 75.6 seconds), which can be attributed to larger number of parameters to train. Also, during testing phase, CNN takes ~$2.5\times$ longer time to generate the output class (0.442 seconds compared to 0.177 seconds).\\
We would like to mention that as the training and testing are done off-line, the training time for the network is not crucial to launch a successful attack, whereas, the number of traces required to recover a key is. An adversary may record all the ciphertexts and corresponding power traces, and later break them off-line. As a result, the high probability of success using Deep Learning techniques for a single-trace attack is of a major concern. 
\section{Conclusion}

This paper presents a practical cross-device attack using deep learning methods even in the presence of misalignment in the captured traces and with significant inter-device variations. Both of these practical issues are challenging to deal with, while implementing a cross-device attack. This article demonstrates how multi-device training improves average test accuracy (e.g., from $61.98\%$ to $91.72\%$ in case of MLP). Moreover, it presents how such attacks can be further improved using pre-processing methods, such as PCA, and DTW, and identifies the best Deep-Learning based approach. PCA based pre-processing improves minimum test accuracy by $10\times$, and DTW-based pre-processing allows subsequent PCA-MLP to maintain its high test accuracy (up to $99.94\%$). Although Deep Learning techniques are more interesting when leakage is harder to model, we note that, success of such techniques lies in data, and pre-processing can help clean the data for more efficient learning.
 
Table IV summarizes the findings by presenting a qualitative comparison of average test accuracy of different methods for both same-device attack and a cross-device attack with multi-device training. In our experiments, CNN maintained a high accuracy in all scenarios, but the best cross-device performance was obtained using the proposed DTW-PCA-MLP. Moreover, compared to the CNN, the proposed DTW-PCA-MLP has a much simpler architecture (less number of tunable parameters), and also requires shorter training time. 
 
Going forward, we would like to study the feasibility of the proposed Deep Learning attack on 32-bit ARM microcontrollers and FPGA-based platforms, which have more widespread use. 

 \newcommand{\noop}[1]{}

\end{document}